\documentclass[12pt,letterpaper]{article}
\def\BibTeX{{\rm B\kern-.05em{\sc i\kern-.025em b}\kern-.08em
    T\kern-.1667em\lower.7ex\hbox{E}\kern-.125emX}}
\usepackage{amssymb, amsmath, amsfonts, epsfig, latexsym, times, setspace}

\def\BibTeX{{\rmfamily B\kern-.05em{\scshape i\kern-.025em b}\kern-.08em \TeX}}

\newtheorem{sideremark}{Remark}
\newcommand{\beq}{\begin{equation}}

\newcommand{\eeq}{\end{equation}}
\def\reals{{\rm I\kern-.17em R}}
\def\nats{{\rm I\kern-.17em N}}

\begin{document}

\title{\Large{\textbf{QoS Provisioning for Multimedia Transmission in Cognitive Radio Networks}}
\thanks{This work is supported by Natural Science and Engineering Research Council of Canada and it is based in part on a paper presented at IEEE WCNC 2009, Budapest, Hungary.}}

\author{\normalsize{F. Richard Yu$^{\dag}$, Bo Sun$^\S$, Vikram Krishnamurthy$^{\ddag}$ and Saqib Ali$^{\dag}$}}

\linespread{1}

\date{\small{$^{\dag}$Department of Systems and Computer Engineering \\ Carleton University, Ottawa, ON, Canada, Email:richard\_yu@carleton.ca \\ Telephone: +1-613-520-2600 ext. 2978, Facsimile: +1-613-520-6623 \\ $^\S$ Department of Computer Science, Lamar University, Beaumont, TX, USA, Email: bsun@my.lamar.edu \\ $^{\ddag}$Department of Electrical and Computer Engineering \\ The University of British Columbia, Vancouver, BC, Canada, Email: vikramk@ece.ubc.ca }}
\maketitle



\noindent {\textbf{\em Abstract - }}
In cognitive radio (CR) networks, the perceived reduction of application layer quality of service (QoS), such as multimedia distortion, by secondary users may impede the success of CR technologies. Most previous work in CR networks ignores application layer QoS. In this paper we take an integrated  design approach to jointly optimize multimedia intra refreshing rate, an application layer parameter, together with access strategy, and spectrum sensing for multimedia transmission in a CR system with time varying wireless channels.  Primary network usage and channel gain are
modeled as a finite state Markov process.  With channel sensing and channel state information errors, the system state cannot be directly observed. We formulate the QoS optimization problem as a partially observable Markov decision process (POMDP).  A low complexity dynamic programming framework is presented to obtain the optimal policy. Simulation results show the effectiveness of the proposed scheme.

\vspace{0.6cm}
\noindent {\textbf{\em Keywords - }} Cognitive radio networks, multimedia, wireless networks.

\vspace{0.6cm}
\noindent

\section{Introduction}

Recent widespread acceptance of wireless
applications has triggered a huge demand for radio
spectrum.  For many years radio spectrum has been assigned to licensed
(primary) users.  Most of the time, some frequency
bands in the radio spectrum remain largely unoccupied by primary
users.  Spectrum usage measurements 
by the Federal Communications Commission (FCC) 
show that at any given time and location, most of the spectrum is actually
idle.  That is, the spectrum shortage results from the spectrum management
policy instead of the actual physical scarcity of usable spectrum.  Cognitive
radio (CR), which has been introduced in \cite{Mit00}, is considered
as an enabling technology that allows unlicensed (secondary) users to operate in the
licensed spectrum bands. This can help overcome the lack of available
spectrum in wireless communications. CR is capable of sensing its
surrounding environment and adapting its internal states by making
corresponding changes in certain operating parameters \cite{Hay05,XZ08,SZ07_1,SZ08}. The
FCC in the United States began to consider more flexible use of
available spectrum. The NeXt Generation program of the Defense
Advanced Research Project Agency also aims to redistribute allocated
spectrum dynamically.

One important application of CR is spectrum overlay dynamic spectrum access (DSA), where secondary users operate in the
licensed band while limiting interference with primary users. Spectrum opportunities are detected and used by secondary users
in the time and frequency domain \cite{ZS07,LYH10,YHT10}. An optimal spectrum sensing strategy is proposed in \cite{ZTS07} to
maximize throughput. A separation principle is established in \cite{CZS08} to decouple the design of the sensing strategy from that of the spectrum sensor and the access strategy. The benefits of cooperation in CR are illustrated in \cite{GL07-I} and
\cite{GL07-II} for two- and multi-user networks,
respectively. A dynamic frequency hopping scheme is presented in \cite{HWV07} for
IEEE 802.22 wireless regional area networks, which is an emerging
standard based on CR technologies. In \cite{JL07}, the authors present a game theoretical dynamic
spectrum sharing framework for analysis of network users' behaviors,
efficient dynamic distributed design, and optimality analysis.  Other
game theoretic DSA methods are presented in \cite{ZT07, EPT07}. The authors in \cite{GTS07} exploit channel availability in
the time domain and demonstrate the throughput performance for a
Bluetooth/WLAN system.  Spectrum opportunity is also exploited in the
time domain in \cite{PSM2005} where the authors present an ad hoc
secondary MAC protocol to facilitate DSA.

Although much work has been done in CR networks, most previous work
considers maximizing the throughput of secondary users as one of the
most important design criteria. As a consequence, other 
QoS measures for secondary users, such as distortion for
multimedia applications, are mostly ignored in the
literature. However, recent work in cross-layer design shows that
maximizing throughput does not necessarily benefit QoS at the application
layer for some multimedia applications, such as video \cite{VS05,  KPS06}. From a user's point of view, QoS at the application layer is
more important than that at other layers. Moreover, CR-based services
for secondary users would have a strictly lower QoS than radio
services that enjoy guaranteed spectrum access \cite{CL07,TZ07_1}. Therefore,
if the application layer QoS is not carefully considered in CR networks, the perceived reduction in QoS associated with CR may impede the success of CR technologies.

Multimedia applications such as video telephony, conferencing, and
video surveillance are being targeted for wireless networks, including
CR networks. Lossy video compression standards, such as MPEG-4 and H.264,
exploit the spatial and temporal redundancy in video streams to reduce the
required bandwidth to transmit video. Compressed video comprises of
intra- and inter-coded frames. The {\em intra refreshing rate} is an
important application layer parameter \cite{HCC02}. Adaptively
adjusting the intra refreshing rate for online video encoding
applications can improve error resilience to the time varying wireless channels available
to secondary users in CR networks.

Cross-layer wireless multimedia transmission, where parameter
optimization is considered jointly across OSI layers,  has been well studied in
the literature \cite{Bou05,YK07,TZ06,TZ05,TZ07_2}. Recent work shows promising improvement to video QoS by considering
resource management, adaptation, and protection strategies available
at the physical, medium access control, and network/transport layers
in conjunction with multimedia compression and streaming algorithms \cite{BHJ02,TZ07_4,YL02,TZ07_3,MYL04,ZS01,ZS04}.
Various channel adaptive distortion driven cross-layer transmission
strategies have been explored. The authors in \cite{VTW06} investigate a classification based system where the
optimal cross-layer strategy for various video and channel conditions
are computed offline thereby reducing the transmission-time complexity
of the compression and transmission strategy.  Within a rate-distortion framework, source coding, retransmission, and adaptive modulation parameters are jointly considered for video summary in \cite{WCW07}.  The authors in
\cite{CMQ07} take a cross-layer approach to allocate power level,
source coding rate, and channel coding rate delivering basic and
enhanced QoS levels for distant and near receivers in a CDMA network. 

Although there are some cross-layer design techniques for wireless multimedia transmission in the literature, little work investigates channel adaptive multimedia transmission over a cognitive radio network. {\em In this paper, we take an integrated design approach to jointly optimize application layer QoS for multimedia transmission over cognitive radio networks.} Based on the sensed channel condition, secondary users can adapt the intra refreshing rate at the application layer, in addition to the parameters at other layers. Some distinct features of the proposed scheme are as follows.
\begin{itemize}
\item For secondary users in CR networks, channel selection for spectrum sensing, access decision, and intra refreshing rate are determined concurrently to maximize the QoS at the application layer (i.e., minimize distortion for video applications).
\item Physical layer channel state information (CSI) (channel gain) is used by secondary users to help make the optimal decision to maximize the application layer QoS.
\item Primary network usage and channel gain are modeled as a finite state Markov process. With channel sensing and CSI errors, the state cannot be directly observed. Following the work in \cite{CZS08}, we formulate the whole system as a partially observable Markov decision process (POMDP) \cite{Cas98}. We extend the scheme to jointly optimizing application layer QoS for multimedia transmission over cognitive radio networks.
\item Using simulation examples, we show that application layer parameters have significant impact on the QoS perceived by secondary users in CR networks. We also show that application layer QoS can be improved significantly if the intra refreshing rate is adapted together with parameters at low layers, such as spectrum sensing. This study reveals a number of interesting observations and provides insights into the design and optimization of CR networks from a cross-layer perspective.
\end{itemize}

The rest of the paper is organized as follows. Section II describes the multimedia transmission over CR networks problem. Section III presents the proposed scheme. Some simulation results are given in Section IV. Finally, we conclude this study in Section V.

 

\section{Multimedia Transmission over Cognitive Radio Networks}

In this section, we describe the multimedia rate-distortion model used in this paper. We then present the system model for multimedia transmission over cognitive radio networks. 

\subsection{Rate-Distortion (R-D) Model for Multimedia Applications}
Wireless channels have limited bandwidth and are error-prone.  Highly
efficient coding algorithms such as H.264 and MPEG-4 can compress
video to reduce the required bandwidth for the video stream.  Rate control is used in video coding to control the video encoder
output bit rate based on various conditions to improve video quality
\cite{LCZ00}. For example, the main tasks of MPEG-4 object-based video
coding are (1) to determine how many bits are assigned to each video
object in the scene and (2) to adjust the quantization parameter to
accurately achieve the target coding bit rate \cite{HKM01}.

Highly compressed video data is vulnerable to packet losses 
where a single bit error may cause severe distortion \cite{SFL00,ZSSK02}.  This vulnerability makes error resilience at the
video encoder essential.  Intra update, also called intra refreshing, of macroblocks (MBs) is one approach for video error resilience and protection \cite{LV00}.  An intra coded MB does not need information from previous frames which may have already been corrupted by channel errors.  This makes intra coded MBs an effective way to mitigate error propagation.  Alternatively, with inter-coded MBs, channel errors from previous frames may still propagate to the current frame along the motion compensation path \cite{CSK00}.

Given a source-coding bit rate $R_s$ and intra refreshing rate, we need a model to estimate the corresponding source distortion
$D_s$.  The authors in \cite{HCC02} provide a closed form distortion model taking into account varying characteristics of the input video, the sophisticated data representation scheme of the coding algorithm, and the intra refreshing rate. Based on the statistical analysis of the error propagation, error concealment, and channel decoding, a theoretical framework is developed to estimate the channel distortion, $D_c$.  Coupled with the R-D model for source coding and time varying wireless channels an adaptive mode selection is proposed for wireless video coding and transmission.

We will use the rate-distortion model described in \cite{HCC02} in our study.  The R-D model facilitates adaptive intra-mode selection
and joint source-channel rate control


The total end-to-end distortion comprises of $D_s$,
the quantization distortion introduced by the lossy video encoder to meet a target
bit rate, and $D_c$, the distortion resulting from
channel errors.  For DCT-based video coding, intra coding of a MB or a frame usually requires more bits than inter
coding since inter coding removes the temporal redundancy between two
neighboring frames.  Let $\beta$ be the intra refreshing rate, the
percentage of MBs coded with intra mode.  Inter coding of MBs has
much better R-D performance than intra mode. Decreasing the intra refreshing
rate decreases the source distortion for a target bit rate.  However
inter coding relies on information in previous frames.  Packet losses due to channel
errors result in error propagation along the motion-compensation path
until the next intra coded MB is received.  Increasing the
intra refreshing rate decreases the channel distortion.  Thus we have a
tradeoff between source and channel distortion when selecting the
intra refreshing rate.  We aim to find
the optimal $\beta$ to minimize the total end-to-end distortion given
the channel bandwidth and packet loss ratio.  

The source distortion is given by
\begin{equation}
D_s(R_s,\beta) = D_s(R_s,0) + \beta(1-\eta + \eta \beta)[D_s(R_s,1) - D_s(R_s,0)],
\end{equation}
where $R_s$ denotes the source coding rate, $\beta$ is the
intra refreshing rate, and $\eta$ is a constant
based on the video sequence.  $D_s(R_s,0)$ and
$D_s(R_s,1)$ denotes the time average all inter- and
intra-mode selection for all frames over $K$ time slots.
\begin{eqnarray}
D_s(R_s,0) = \frac{1}{K} \sum_{k=1}^{K} \frac{1}{M_k} \sum_{m=1}^{M_k} D_s(Rs,0,m),
\end{eqnarray}
\begin{eqnarray}
D_s(R_s,1) = \frac{1}{K} \sum_{k=1}^{K} \frac{1}{M_k} \sum_{m=1}^{M_k} D_s(Rs,1,m),
\end{eqnarray}
where $M_k$ is the number of inter/intra frames in time slot $k$. The average channel distortion for each time slot is given by
\begin{eqnarray}
D_c(p,\beta) = \left( \frac{a}{1-b+b \beta} \right) \left( \frac{p}{1-p} \right) E[F_d(m,m-1)],
\end{eqnarray}
where $p$ is the packet loss rate, $b$ is a constant describing motion
randomness of the video scene, $a$ is the energy loss ratio of the
encoder filter, and $E[F_d(m,m-1)]$ is the average value of the frame difference
$F_d(m,m-1)$ over $K$ slots. We will use the same error concealment
strategy and packet loss ratio derivation as described in \cite{HCC02}.

The total average distortion is given by
\begin{eqnarray}
D(R_s, p, \beta) = D_s(R_s,\beta) + D_c(p,\beta).
\end{eqnarray}
The optimum $\beta^*$ is then selected to minimize the total distortion.
\begin{eqnarray}
\beta^* = \hbox{arg} \min_\beta D(R_s,p, \beta).
\end{eqnarray}

\subsection{System Model}
Consider a spectrum that consists of $N$ channels, each with bandwidth
$W(n)$, $1\le n \le N$.  These $N$ channels are licensed to primary users.  Time is divided into slots of equal length $T$.  Slot $k$ refers to the discrete time period $[kT, (k+1)T]$.    

When the slot is not in use by primary users, it will be comprised of
AWGN noise and fading.  The fading process and primary usage for a
channel can be represented by a stationary and ergodic $S$-state
Markov chain.  Let $i$ and $\gamma$ denote the instantaneous channel
state and fading gain, respectively.  When the channel is in state
$i$, the quantized fading gain is $\gamma_i$, where $\gamma_{i} \le \gamma \le \gamma_{i+1}$, $1 \le i \le S-1$. When the channel is in state $i=S$, the channel is in use by the primary network.  We
assume that the phase of the channel attenuation can be perfectly
estimated and removed at the receiver.  The $S$-state Markov channel
model is completely described by its stationary distribution of each
channel state $i$, denoted by $p(i)$, and the probability of
transitioning from state $i$ into state $j$ after each time slot, denoted by $\{P_{i,j}\}$, $1 \le i,j \le S$.  

In general, a finite state Markov channel (FSMC) model is constructed for a
particular fading distribution by first partitioning the range of the
fading gain into a finite number of sections.  Then each section of
the gain value corresponds to a state in the Markov chain.  The application of
FSMC to model Rayleigh channels has been well studied in \cite{TB00,BMG02}.  Given
knowledge of the fading process and primary network usage, the
stationary distribution $p(i)$ as well as channel state transition probabilities $\{P_{i,j}\}$ can be derived. Once a channel gain has been determined for states 1,2,..., $S$-1, the packet loss ratio is determined for each state based on the
modulation and channel coding schemes.  The intra refreshing rate that minimizes the total distortion for each state can then be calculated using the Rate-Distortion model.

\begin{figure}
\begin{center}
\vspace{-5cm}
\epsfig{figure=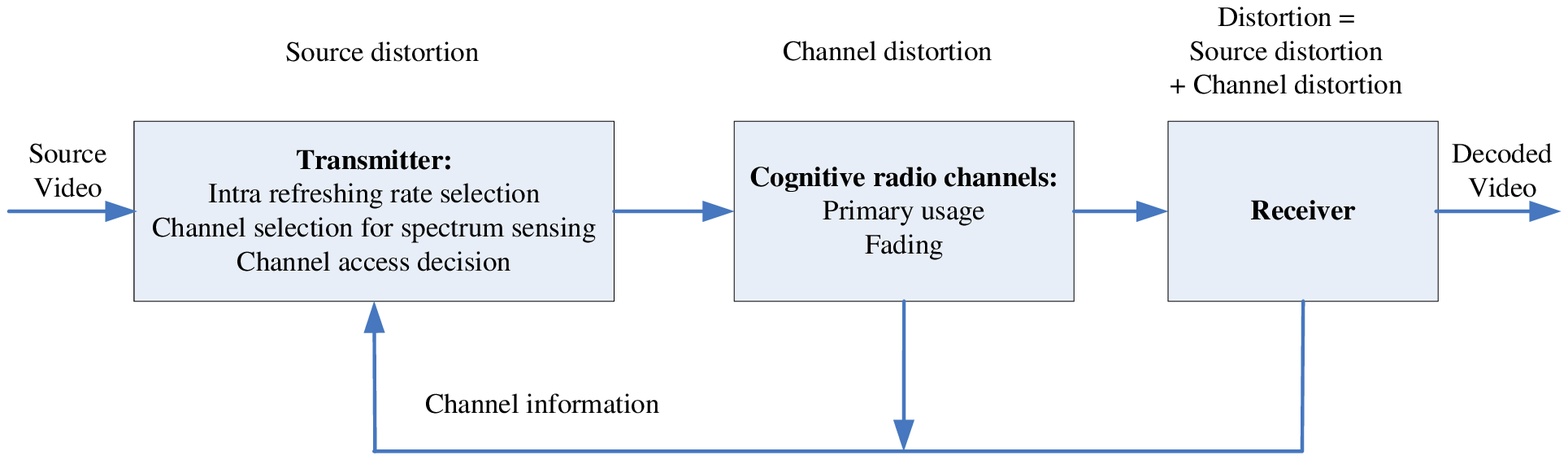, width=1\linewidth}
\vspace{-13cm}
\caption{The Block diagram of multimedia transmission over cognitive radio networks.}
\label{figure:system_block}
\end{center}
\end{figure}

At the beginning of a slot, the transmitter of secondary users will select a set of channels to sense.  Based on the
sensing outcome, the transmitter will decide whether or not to access a channel.
If the transmitter decides to access a channel, some application layer parameters will be selected and the video content will be transmitted.  At the end of the slot, the receiver will
acknowledge the transfer by sending the perceived channel gain back
to the transmitter.  We will assume a system for real-time multimedia applications where packets
are discarded if a primary user is using the slot or if the channel is
not accessed.  The system block diagram
showing video transmission between two secondary users is shown in
Fig. \ref{figure:system_block}.

\section{Solving the Multimedia Transmission over Cognitive Radio Networks Problem}
In CR networks with multimedia applications, we need to determine the optimal policy for channel sensing selection, sensor operating
point, access decision, and intra refreshing rate to minimize application layer distortion subject to the system probability of collision. With channel sensing and CSI errors, the system state cannot be directly observed. Following the work in \cite{CZS08}, we formulate the whole system as a partially observable Markov decision process (POMDP). Deriving a single POMDP formulation for all policies under the probability of collision constraint would result in a constrained
POMDP.  However, constrained POMDPs require randomized policies to achieve
optimality, which is often intractable. Therefore, we use the separation principle in \cite{CZS08} for the sensor operating point and the access decision.  The spectrum sensor operating point is set such that $\delta = \zeta$, where $\delta$ is the probability of miss detection of the busy channel used by primary users and $\zeta$ is the required probability of collision.

At the beginning of the slot, the system transitions to a new state. Using a POMDP derived policy, a channel is selected for spectrum sensing. An access decision is then made based on the sensing observation.  Using the belief of the channel state, an intra refreshing rate is selected.   The receiver acknowledges the transfer by sending the quantized perceived channel gain back to the secondary transmitter.  The immediate cost for the time slot is derived based on the previous operations in the slot.

The system can be formulated as a POMDP with states, actions, transition probabilities, observations, and cost
structures as follows.  

\subsection{State Space, Transition Probabilities and Observation Space} 

The system state is given by the network usage of primary users and channel state information.  Let $\{X(n)\}$ denote an $S$-state Markov chain for channel $n$, $X(n) \in \mathbb{X} = \{e_1, e_2, ..., e_{S-1},e_S\}$, where $e_i$ denotes the $S$-dimensional unit vector with 1 in the $i$th position and zeros elsewhere. The system with $N$ channels is modeled as a discrete-time homogeneous
Markov process with $S^N$ states.  The system state in time slot $k$ is given by $V_k = [X_k(1), ..., X_k(N)]$.  

To simplify the presentation, we consider a system with a single
channel in the formulation. It is straightforward to extend the
formulation to include multiple channels which is considered in our simulations. For the system with a single channel, $V_k = X_k$. The transition probabilities of the system state are given by the $S \times S$ matrix $A$. We assume the transition probabilities are known based on network usage and channel fading characteristics.

The observation available to the secondary transmitter and receiver is the sensed channel and channel gain acknowledgment, $Y_k \in \mathbb{Y}$, where $\mathbb{Y} = \{{\gamma}_1, ..., {\gamma}_{S-1}, {\gamma}_S \mbox{ (The channel is used by primary users)}\}$ and ${\gamma}_i < {\gamma}_j, \forall i < j$.

The spectrum sensor observation may be different at the
transmitter and receiver.  If the transmitter and receiver use the same observations
to derive the information state (described in the following Subsection), then the information state can be used to
maintain frequency hopping synchronization.  Thus the information
state will be updated with $Y_k$ and will not include the spectrum
sensor observation.  

Let $B(y, x, a) = \mbox{Pr}\{y | x, a\}$ denote the conditional probability of observing $y$ given that the system state is in state $x$ and composite action $a$ was taken.

\begin{equation}
B(y,x,a) = \left\{ 
\begin{array}{ll}
P_{ce}(x, v(y)) (1-\epsilon), & \mbox{if } y \neq {\gamma}_S, x \neq e_S,\\
\epsilon, & \mbox{if } y = {\gamma}_S, x \neq e_S\\
0, & \mbox{if } y \neq {\gamma}_S, x = e_S\\
1, & \mbox{if } y = {\gamma}_S, x = e_S
\end{array}, \right.
\end{equation}
where $\epsilon$ is the probability of miss detection of the idle channel and $v(y) = i$, $1 < i < S$ given $y = {\gamma}_i$. When the channel is available and accessed, the probability of channel estimate by the receiver is given by $P_{ce}(x, v(y))$ .

Using the work from Hoang and Motani \cite{HM04}, we assume the channel estimation error has a Gaussian distribution with zero mean and $\sigma^2$ variance.  At a particular time and channel, the estimated channel gain is
\begin{equation}
\hat{\gamma} = \gamma_i + w,
\end{equation}
where $\gamma_i$ is the actual channel gain and $w$ is a Gaussian
random variable with zero mean and $\sigma^2$ variance.  The receiver
then quantizes the channel gain to the nearest possible value.  The
probability that $\hat{\gamma}$ is closest to $\gamma_j$ is given by
\begin{equation}
P_{ce}(i,j) = \left\{
\begin{array}{ll}
\frac{1}{2} \left[ \mbox{erf} \left(\frac{\gamma_j + \gamma_{j+1} - 2\gamma_i}{2\sqrt{2}\sigma} \right) - \mbox{erf} \left(\frac {\gamma_j + \gamma_{j-1} - 2\gamma_i}{2\sqrt{2}\sigma}\right) \right], & \mbox{if } j \neq e_1, e_{S-1}, e_S \\
\frac{1}{2} \left[1 + \mbox{erf} \left(\frac{\gamma_1 + \gamma_2 - 2\gamma_i}{2\sqrt{2}\sigma}\right) \right], & \mbox{if } j = e_1 \\
\frac{1}{2} \left[1 - \mbox{erf} \left(\frac{\gamma_{S-2} + \gamma_{S-1} - 2\gamma_i}{2\sqrt{2}\sigma}\right)\right], & \mbox{if } j = e_{S-1} \\
0, & \mbox{if } j = e_S 
\end{array}, \right.
\end{equation}
where $\mbox{erf}()$ denotes the error function.

\subsection{Information State} 
\begin{figure}
\begin{center}
\epsfig{figure=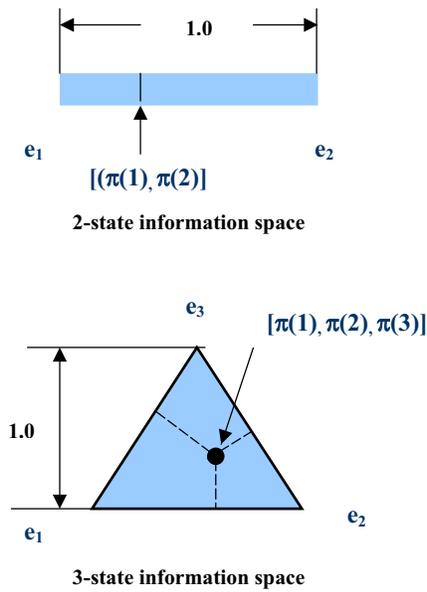, width=0.35\linewidth}
\end{center}
\caption{Information state in POMDP.}
\label{InfoState}
\end{figure}

Information state is an important concept in POMDP. We will refer to a probability distribution over states as the information state and the entire probability space (the set of all possible probability distributions) as the information space. The information spaces for 2-state and 3-state systems are shown in Fig.~\ref{InfoState}. For a system with two states, its information space is a one-dimension
line. The distance from the right end is the first component $\pi(1)$ and the distance from the left end is the second component $ \pi(2)$. For the system with 3 states, its information space is a two-dimension triangle. The value of a point in the information space can be obtained from the perpendicular distance from the sides of the triangle. An information state is a sufficient statistic for the decision and observation history.

\subsection{Action Space} 
Due to hardware limitations, we will assume that a secondary user is
equipped with a single Neyman-Pearson energy detector and can only
sense $L=1$ channel at each time instant. In each slot $k$, the
secondary user needs to decide whether or not to sense, determine which sensor operating
point on the Receiver Operating Curve (ROC) curve to use, whether to access the channel, and
which quantized intra refreshing rate to use.  Thus the action space consists of four parts: a channel selection decision $a_s(k) \in \{0
(\mbox{no sense}), 1 (\mbox{sense})\}$, a spectrum sensor design $(\epsilon(k),
\delta(k)) \in \mathbb{A}_{\epsilon \delta}$ where $\mathbb{A}_{\epsilon \delta}$ are valid
points on the ROC curve, an access decision $a_a(k) \in \{0(\mbox{no access}), 1 (\mbox{access})\}$, and an intra refreshing rate $\beta(k) \in \mathbb{A}_{\beta}$.  The composite action in slot $k$ is denoted by $a_k = \{a_s(k), (\epsilon(k), \delta(k)), a_a(k), \beta(k) \} \in (\{0,1\}, \mathbb{A}_{\epsilon \delta}, \{0,1\}, \mathbb{A}_\beta)$.

Due to sensing and channel estimation errors, a secondary user cannot directly observe the true system state.  It can
infer the system state from its decision and observation history encapsulated
by the information state.  Information state $\pi_k = \{\lambda_x(k)\}_{x \in \mathbb{X}} \in \Pi(\mathbb{X})$ where $\lambda_x(k) \in [0,1]$
denotes the conditional probability (given decision and observation
history) that the system state is in $x \in \mathbb{X}$ at the beginning of slot $k$
prior to state transition. $\Pi(\mathbb{X}) = \{\lambda_x(k) \in
[0,1], \sum_{x \in \mathbb{X}} \lambda_x = 1\}$ denotes the information
space that includes all possible probability mass functions on the
state space $\mathbb{X}$.  

At the end of the time slot, the transmitter receives observation $Y_k$.  The
information state is then updated using Bayes' rule {\em before} state transition

\begin{eqnarray}
\lambda_{k+1} = \frac{\sum_{x' \in \mathbb{X}} \lambda_{x'}(k) A_{x',x}
  B(y_k, x_k, a_k)}{\sum_{x \in \mathbb{X}} \sum_{x' \in \mathbb{X}} \lambda_{x'}(k) A_{x',x} B(y_k, x_k, a_k)}.
\label{IS_Update}
\end{eqnarray}

Given information vector $\pi_k$ the
distribution of the system state $X_k$ in slot $k$ {\em after} state
transition is then given by
\begin{eqnarray}
\mbox{Pr}\{X_k = x\} = \sum_{x' \in \mathbb{X}} \lambda_{x'}(k)A_{x',x}
\; \forall x \in \mathbb{X}.
\end{eqnarray}

\subsection{Cost and Policy} 
From a user's point of view, QoS at application layer is more important than at other layers. Therefore, we model multimedia distortion as the immediate cost in our scheme. The immediate cost in time slot $k$ is defined as
\begin{eqnarray}
C_k = D(R, p(x_k, a_k), \beta(k)),
\end{eqnarray}
where $R$ is the target bit rate and $p(x_k, a_k)$ denotes the packet
loss ratio when the system is in state $x_k$ and composite action
$a_k$ is taken in time slot $k$.  We assume $a_a(k) = 0 (\hbox{no
  access})$ is the equivalent to 100\% packet loss.  

The expected total cost of the POMDP represents the overall distortion for a video sequence transmitted over $K$ slots and can be expressed as

\begin{eqnarray}
J_\mu = \mathbb{E}_{\{\mu_s, \mu_{\epsilon \delta}, \mu_a, \mu_{\beta}\}} \left[\sum_{k=1}^{K} D(R, p(x_k, a_k), \beta(k))\right],
\end{eqnarray}
where $\mathbb{E}_{\{\mu_s, \mu_{\epsilon \delta}, \mu_a, \mu_{\beta}\}}$ indicates the expectation given that policies $\mu_s, \mu_{\epsilon \delta}, \mu_a, \mu_{\beta}$ are employed.  

A channel sensing policy $\mu_s$ specifies a channel to sense, $a_s$.  A sensor operating policy $\mu_{\epsilon \delta}$ specifies a spectrum sensor design $(\epsilon, \delta) \in \mathbb{A}_{\epsilon \delta}$ based on the system tolerable probability of collision, $\zeta$.  An access policy $\mu_a$ specifies the access decision $a_a \in \{0,1\}$.  An intra refreshing policy $\mu_\beta$ specifies the intra refreshing decision ${\beta} \in \mathbb{A}_{\beta}$ based on the current information state $\pi_k$. 

\subsection{Objective and Constraint}
We aim to develop the joint design of an optimal policy for multimedia transmission over CR networks, $\{\mu_s^*, \mu_{\epsilon \delta}^*, \mu_a^*, \mu_{\beta}^* \}$, that minimizes the expected total distortion in $K$ slots under the collision constraint $P_c$.
\begin{eqnarray}
\{\mu_s^*, \mu_{\epsilon \delta}^*, \mu_a^*, \mu_{\beta}^*\} & = & \hbox{arg} \min_{\mu_s, \mu_{\epsilon \delta}, \mu_a, \mu_{\beta}}
\mathbb{E}_{\{\mu_s, \mu_{\epsilon \delta}, \mu_a, \mu_{\beta}\}} \left[\sum_{k=1}^{K} D(R, p(x_k, a_k), \beta(k))\right] \\
\hbox{subject to} \nonumber \\
P_c(k) & = & \hbox{Pr}\{a_a(k) = 1 | X_k = e_S\} < \zeta \;, \forall k \in K.
\end{eqnarray}

\subsection{Value Function}
Let $J_k(\pi)$ be the value function that represents the minimum expected cost that can be obtained starting from slot $k$ $(1
\le k \le K)$ given information state $\pi_k$ at the beginning of
slot $k$.  Given that the secondary user takes action $a_k$ and
observes acknowledgment $Y_k = y_k$, the cost that
can be accumulated starting from slot $k$ consists of the immediate
cost $C_k = D(R, p(x_k, a_k), \beta(k))$ and the minimum expected future
cost $J_{k+1}(\pi+1)$. $\pi_{k+1} = \{ \lambda_x(k+1) \}_{x \in \mathbb{X}} = U(\pi_k | a_k, y_k)$, which represents the updated knowledge of system state after incorporating the action $a_k$ and the acknowledgment $y_k$ in slot $k$.  The sensing policy is then given by
\begin{eqnarray}
J_k(\pi_k) &=& \min_{a \in \mathbb{A}} \sum_{x \in \mathbb{X}}
\sum_{x' \in \mathbb{X}} \lambda_{x'}(k) A_{x',x} 
 \sum_{j=e_1}^{e_S} B(y_k, j, a_k) [D(R, p(x_k, a_k), \beta(k)) \nonumber \\
 && +  J_{k+1}(U(\pi_k | a_k, y_k))] , 1 \le k \le K-1  \label{Value_Function} \\
J_K(\pi_K) & = & \min_{a \in \mathbb{A}} \sum_{x \in \mathbb{X}}
\sum_{x' \in \mathbb{X}} \lambda_{x'}(K) A_{x',x}
 \left[ \sum_{j=e_1}^{e_S}
B(y_K, j, a_K) D(R, p(x_K, a_K), \beta(K)) \right].
\end{eqnarray}

The value function of an unconstrained POMDP with finite action space is piecewise-linear convex and can be solved using linear programming techniques \cite{Lov91}. An excellent overview of computationally efficient algorithms are given in \cite{Cas98} and can be used to solve for the optimum sensing policy. In general, the number of linear segments that characterize the value function can grow exponentially. In 1991, Lovejoy proposed an ingenious suboptimal algorithm for POMDPs \cite{Lov91b}. Based on Lovejoy's algorithm, the value function can be upper and lower bounded and efficient suboptimal solutions can be developed as in Subsection V-D of \cite{Kri05b}. By considering only a subset of the piecewise linear segments that characterize the value function and discarding the other segments, one can reduce the computational complexity. Due to the space limitation, please refer to Subsection V-D of \cite{Kri05b} for details. Moreover, solving the POMDP can be done off-line during system initialization. During the real-time multimedia transmission, a node just needs to find the value for specific information state according to (\ref{Value_Function}) and update the information state according to (\ref{IS_Update}), which introduces little computational complexity. Finally, by imposing structural assumptions on the transition probabilities, cost and observation probabilities, one can prove in some cases that the optimal policy is a threshold policy \cite{KD07}.

\subsection{Intra Refreshing Strategy}
For a selected channel, the optimum $\beta$ selected
corresponds to the most likely available state based on $\pi_k$. Due to the asymptotic nature of the channel distortion, a busy or
unaccessed channel has infinite distortion.  In this case, 
$\beta$ has no influence on the total distortion.  If the most likely state based
on $\pi_k$ corresponds to a busy state then the optimum $\beta$ is to
select a $\beta$ corresponding to the most likely available state.  That way if the
information state suggests the channel is busy but in reality it is
available then a $\beta$ has been selected that will minimize the
effect of this error.

\section{Simulation Results and Discussions}
In order to evaluate the performance of our proposed scheme, we have carried out a set of simulation experiments using the ns-2 simulator. All simulations were run  on a computer equipped with Window 7, Intel Core 2 Duo P8400 CPU (2.26 Ghz) and 4GB memory. The choice for the total time slot number $K$ in the dynamic programming depends on the convergence rate of the POMDP program. State transition probabilities, observation probabilities and value functions have effects on the convergence rate \cite{Cas98}, \cite{Lov91}. In our simulations, the POMDP program was run over a horizon of $K = 200$. It is reasonable to use $K = 200$ to approximate the problem with infinite horizon. We first consider a system with one channel in Subsections 4.1 - 4.3. Then, we consider a system with two channels in Subsections 4.4 - 4.5. In all figures the curves represent the average values, while the error bars represent the confidence intervals for 95 percent confidence for 50 different instances (seeds). 

We consider the system performance in the following four cases: (1) using perfect knowledge of the system thus making optimal decisions, which is the best case possible, (2) making decisions based on the most likely state indicated by the information state, which is our proposed scheme, (3) making decisions solely based on the channel gain provided in the last acknowledgment, and (4) using a constant $\beta$, which represents existing schemes that do
not consider application layer QoS. Our goal is to compare the distortion of different schemes as opposed to determining the absolute distortion. We use an average distortion metric that refers to the
average distortion over the time slots when the channel is available
and accessed. Video rate-distortion parameters remain constant for the duration of the simulation.  The
same distortion parameters are used for all simulations.  $D_s(R_s,0)
= 74$.  $D_s(R_s,1) = 124$.  $\eta = 1.4$.  $a = 0.01$.  $b =
1.0$.  $E[F_d(m,m-1)] = 100$.  

\subsection{Performance Improvement}
\begin{figure}[!t]
\centering
\hspace{0.5cm}
\epsfig{figure=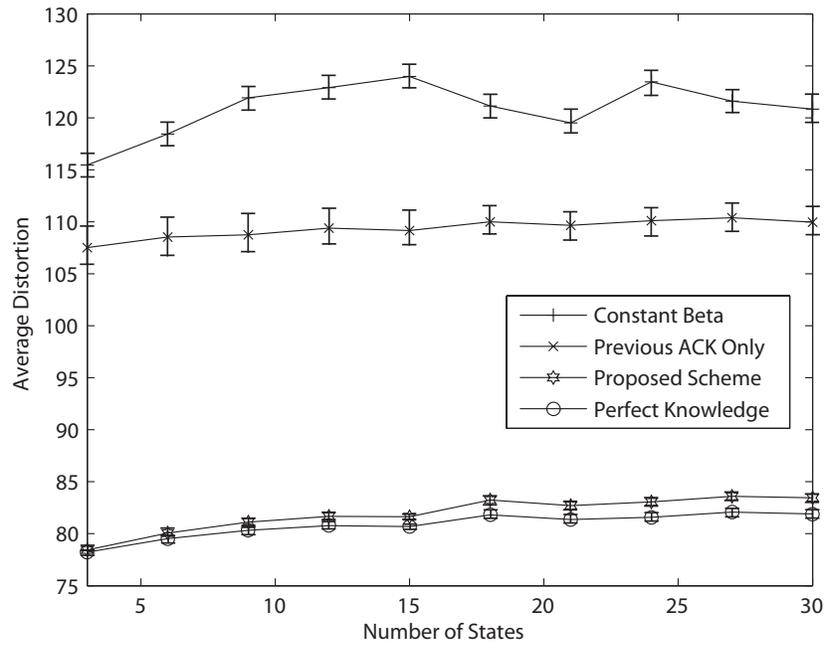, width=0.7\linewidth}
\caption{Average distortion vs. the number of states in different schemes.}
\label{fig_multi_state}
\end{figure}

Fig. \ref{fig_multi_state} shows the distortion of different schemes. The number of states refers to $S-1$ quantized channel gains and one busy channel state. For simplicity we derive a transition matrix based on the probability that any available state stays in the same
state, $\mbox{Pr}\{X_{k+1} = v | X_k = v\}$, the probability of
transitioning from an available state to a busy state,
$\mbox{Pr}\{X_{k+1} = z | X_k = v\}$, and the probability of a busy state
staying busy, $\mbox{Pr}\{X_{k+1} = z | X_k = z\}$, $\forall v \in \{e_1, e_2, ..., e_{S-1}\}, z = e_S$, where $v$ and $z$
indicate available and busy states, respectively. The following parameter values are used in this example. $\mbox{Pr}\{X_{k+1} = v | X_k = v\} = 0.85$, $\mbox{Pr}\{X_{k+1} = z |
X_k = v\} = 0.05$, $\mbox{Pr}\{X_{k+1} = z | X_k = z\} = 0.1$,
$\epsilon = 0.6$, $\sigma = 0.1$.  From Fig. \ref{fig_multi_state}, we can see that when perfect knowledge of the
channel state is available, perfect decisions can be made for
each time slot thus method (1) has the lowest average distortion.
The more realistic cases occur in the presence of sensing and CSI
errors.  Our proposed method (i.e., method 2) uses the information state
to select the most likely optimal decisions.  This method tracks the
ideal case fairly closely.  Both method 3 and method 4 have worse performance compared to the proposed scheme. This illustrates the performance improvement of the proposed scheme over existing schemes. In addition, we also notice that using a constant $\beta$ (i.e., method 4) can be worse than making decisions based solely on the previous acknowledgment (i.e., method 3), which shows the need to consider application layer parameters and application QoS. Moreover, increasing the number of channel states changes the characteristics of the channel. Consequently, the likelihood that the underlying system is in a state where the constant $\beta$ is optimal decreases. Therefore, the performance of using the constant $\beta$ is not stable with the increasing of the number of states. The application layer parameter, $\beta$, should be adapted together with parameters at low layers. 

\subsection{Effects of the Parameters in the State Transition Matrix}
\begin{figure}[!t]
\hspace{3cm}
\epsfig{figure=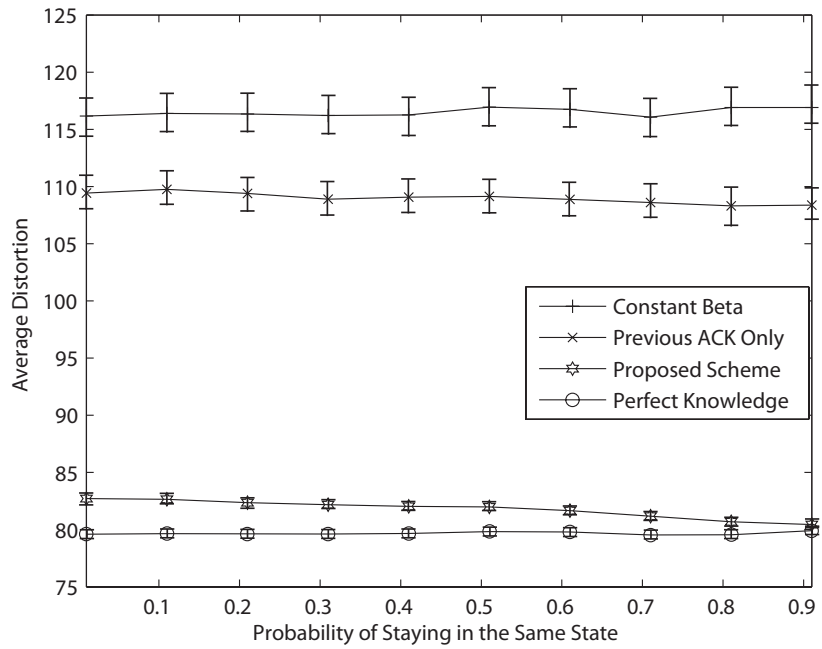, width=0.7\linewidth}
\caption{Average distortion vs. the probability of staying in the same state.}
\label{fig_prob_same}
\end{figure}

\begin{figure}[!t]
\hspace{3cm}
\epsfig{figure=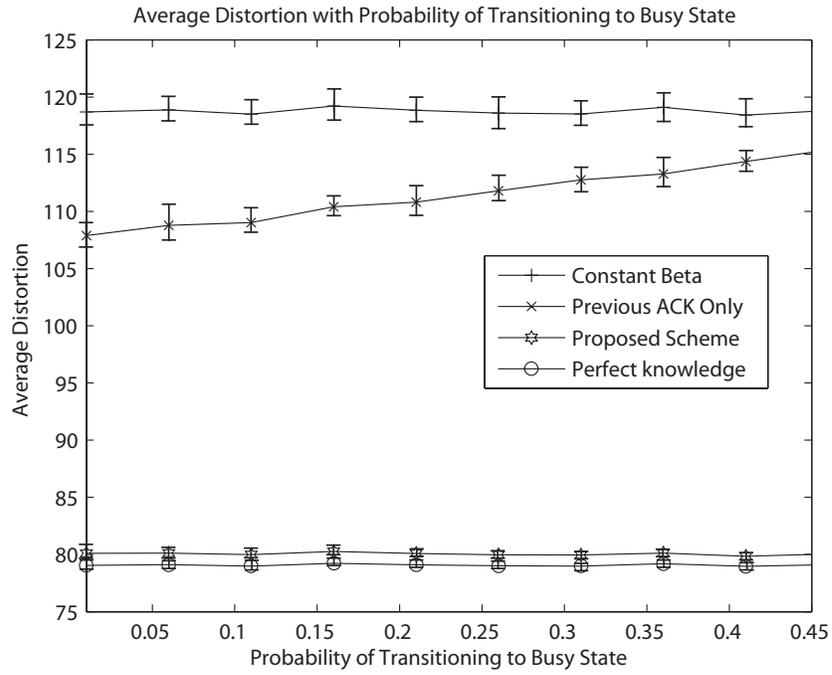, width=0.7\linewidth}
\caption{Average distortion vs. the probability of transitioning to the busy state. }
\label{fig_pri_usage}
\end{figure}

We evaluate how the parameters in the transition matrix affect the average distortion.  The transition
matrix can be selected based on channel fading and primary usage. We ignore quantization errors caused
by the limited number of states and assume the actual channel gain matches
the state channel gain. Fig. \ref{fig_prob_same} and Fig. \ref{fig_pri_usage} show the
simulation results across $\mbox{Pr}\{X_{k+1} = v | X_k = v\}$ and
$\mbox{Pr}\{X_{k+1} = z | X_k = v\}$, respectively. In Fig. \ref{fig_prob_same}, there are 5 states. $\epsilon=0.6$. $\mbox{Pr}\{X_{k+1} = z | X_k =
  v\}=0.05$. This example demonstrates the cognitive nature of the system. Our proposed method
(i.e., method 2) approaches the method of using perfect knowledge of
the channel state as $\mbox{Pr}\{X_{k+1} = v | X_k = v\}$ approaches
1.  That is, the performance improves as the system dynamics slows down since it is easier to predict the
actual system state.  5 states are used in Fig. \ref{fig_pri_usage}. $\epsilon=0.6$. $\mbox{Pr}\{X_{k+1} = v | X_k = v\}=0.50$. From this figure, we can see that $\mbox{Pr}\{X_{k+1} = z | X_k = v\}$ has
little impact to the performance of the proposed method. The reason for this observation is that increasing  $\mbox{Pr}\{X_{k+1}
= z | X_k = v\}$ will increase the likelihood the system transitions to the busy state, which has little affect on the average distortion when
the channel is available and accessed.  

\subsection{Effects of the Parameters in the Observation Matrix}
\begin{figure}[!t]
\hspace{3cm}
\epsfig{figure=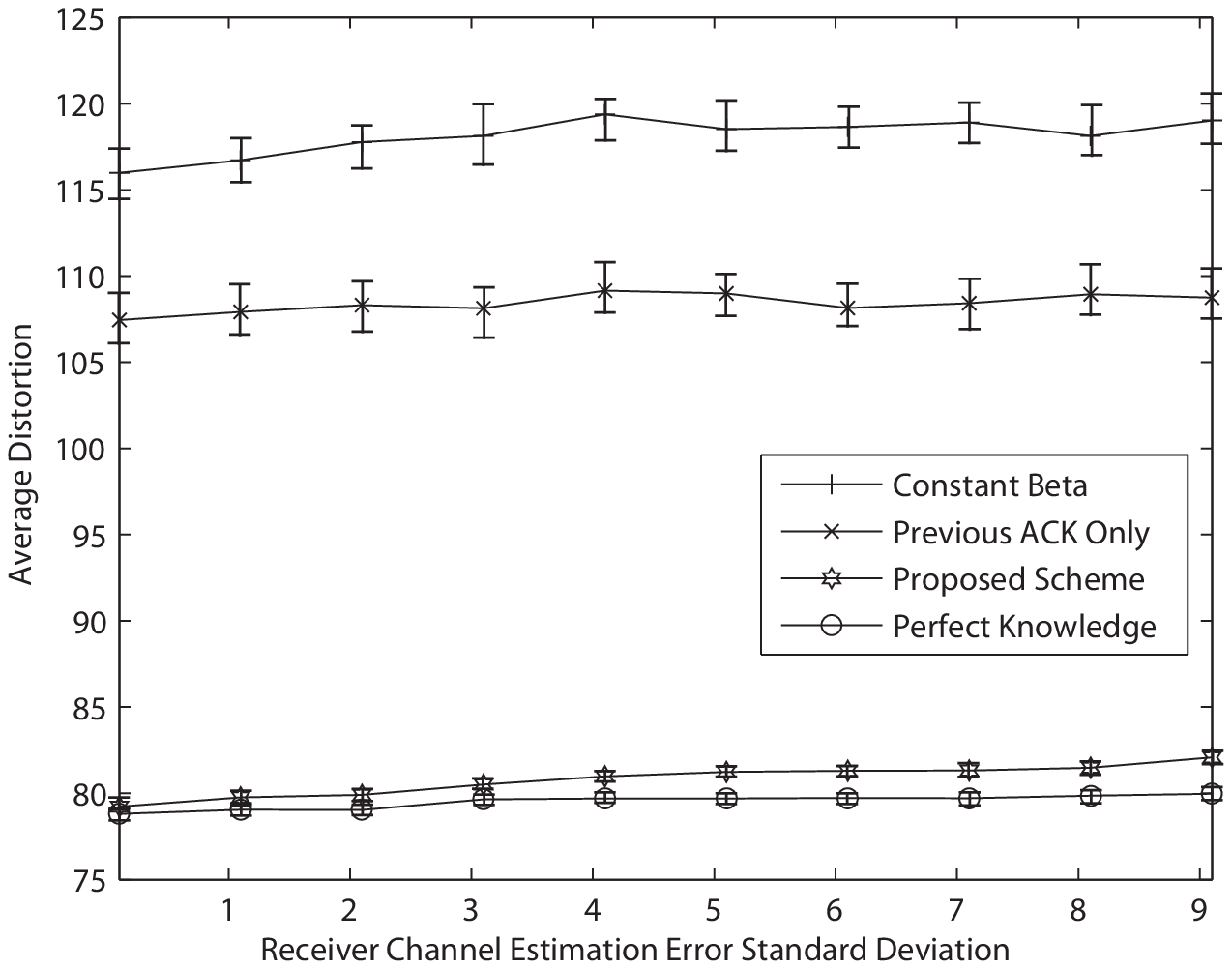, width=0.7\linewidth}
\caption{Average distortion vs. the receiver channel estimation standard deviation, $\sigma$.}
\label{fig_std_dev}
\end{figure}

\begin{figure}[!t]
\hspace{3cm}
\epsfig{figure=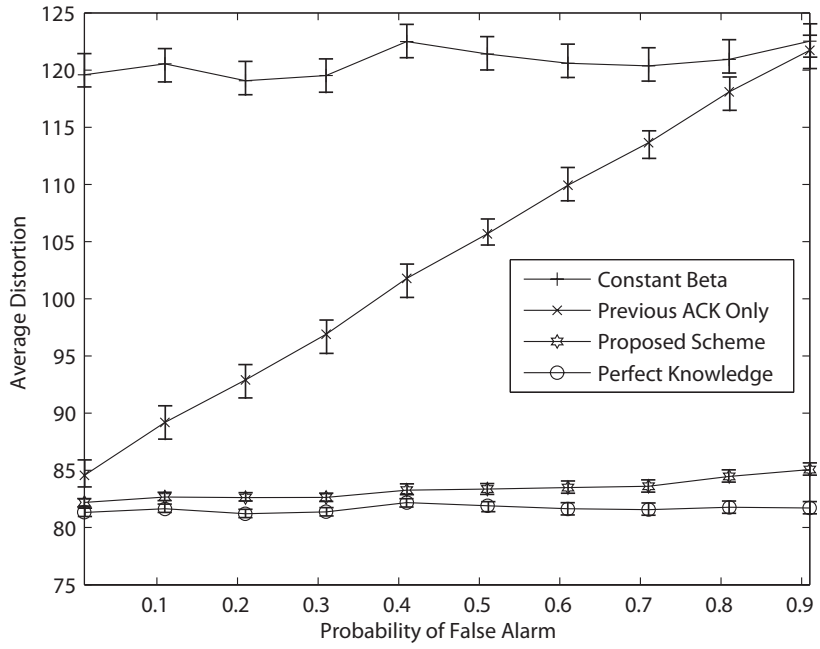, width=0.7\linewidth}
\caption{Average distortion vs. the sensor operating point, $\epsilon$.}
\label{fig_epsilon}
\end{figure}

The observation matrix is derived from the sensor
operating point, $\epsilon$, and the standard deviation of the
receiver channel estimation error, $\sigma$. Fig. \ref{fig_std_dev}
and Fig. \ref{fig_epsilon} show how $\sigma$ and $\epsilon$ affect the
average 
distortion. The following parameters are used in Fig. \ref{fig_std_dev}. There are 5 states, $\epsilon=0.6$, $\mbox{Pr}\{X_{k+1} = v | X_k =
  v\}=0.85$, and $\mbox{Pr}\{X_{k+1} = z | X_k = v\}=0.05$. We can see from Fig. \ref{fig_std_dev}, as the
receiver estimation degrades, the acknowledgment provides less
information on the actual channel gain and the average distortion of
our method increases.  $\epsilon$ and $\delta$ are
related based on the sensor ROC, and
adjusting $\epsilon$ implies a change to the system probability of
collision requirement. In Fig. \ref{fig_epsilon}, $\mbox{Pr}\{X_{k+1}
= v | X_k = v\}=0.85$, $\mbox{Pr}\{X_{k+1} = z | X_k = v\}=0.05$, and
$\sigma = 0.1$. This figure shows that the average distortion
increases as the probability of false alarm increases.

\subsection{Effects of the Transition Matrix on Channel Selection Policy}
We consider a system with $N=2$ channels and $S=3$ states to evaluate the
performance of the channel selection policy.  We will use a spectrum
utilization (SU) metric to evaluate the sensor policy performance.
SU represents the percentage of time slots where an
available channel was selected for sensing. SU is an important parameter when evaluating video
QoS.  The channel distortion is infinite when a channel is busy or not
accessed.  Improving the SU will reduce the percentage of time slots
where a busy channel was selected for sensing thus improving the
application layer QoS.  The application layer QoS is improved using a
two step process.  First we select a channel to maximize SU thus reducing
the large distortion introduced when the channel is unavailable.
Second for an available and accessed channel, we select the
intra refreshing rate to minimize distortion for a particular channel
gain.  

The two channels, channel $1$ and channel $2$, are simulated having
the same number of states (i.e. quantized channel gains) and
observation probabilities but asymmetric transition probabilities.
Channel $2$ will have a higher primary usage than channel $1$.
Based on previous observations, actions, and the POMDP derived policy
the secondary transmitter/receiver pair dynamically selects the channel that will most
likely maximize application layer QoS.  

We evaluate SU and average distortion performance for three cases (1) POMDP channel
selection, which is our proposed scheme, (2) randomly selecting channel
$1$ or $2$ and using a constant $\beta=0.1$, which represents a
non-adaptive scheme, and (3) using perfect knowledge of
the system state, which represents the ideal case.

\begin{figure}[!t]
\hspace{3cm}
\epsfig{figure=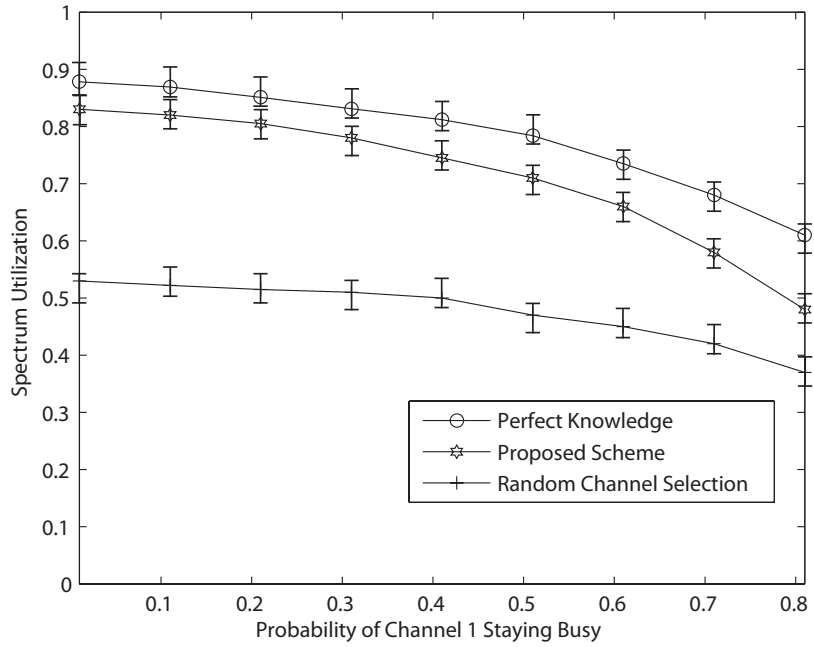, width=0.7\linewidth}
\caption{Two Channel Scenario: Spectrum utilization vs. the probability of staying in the busy state of channel 1.}
\label{fig_2c_spec_stay_busy}
\end{figure}

\begin{figure}[!t]
\hspace{3cm}
\epsfig{figure=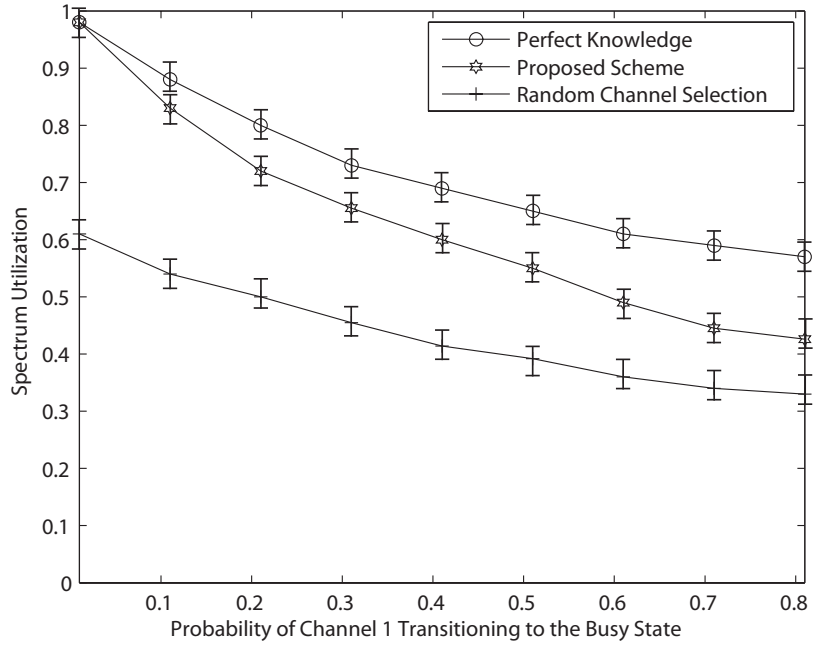, width=0.7\linewidth}
\caption{Two Channel Scenario: Spectrum utilization vs. the probability of transitioning to the busy state of channel 1.}
\label{fig_2c_spec_trans_busy}
\end{figure}

SU performance with varying transition matrix parameters is shown in
Fig. \ref{fig_2c_spec_stay_busy} and Fig. \ref{fig_2c_spec_trans_busy}.  In
both plots we only vary the
transition matrix parameters of channel $1$.  Both channels
have equal observation matrix parameters $\epsilon^1=\epsilon^2=0.62$ and
$\sigma^1=\sigma^2=0.1$.  

In Fig. \ref{fig_2c_spec_stay_busy} we vary the probability channel $1$
stays busy, $\mbox{Pr}\{X_{k+1}^1 = z | X_k^1 = z\}$.
$\mbox{Pr}\{X_{k+1}^1 = z | X_k^1 = v\} = 0.2$. $\mbox{Pr}\{X_{k+1}^2
= z | X_k^2 = z\} = 0.8$. $\mbox{Pr}\{X_{k+1}^2 = z | X_k^2 = v\} =
0.6$.  In Fig.  \ref{fig_2c_spec_trans_busy} we vary the probability
channel $1$ transitions to the
busy state, $\mbox{Pr}\{X_{k+1}^1 = z | X_k^1 = v\}$.  $\mbox{Pr}\{X_{k+1}^1 = z | X_k^1 = z\} = 0.4$.
$\mbox{Pr}\{X_{k+1}^2 = z | X_k^2 = z\} = 0.8$. $\mbox{Pr}\{X_{k+1}^2 = z | X_k^2 = v\} =
0.6$.  

\begin{figure}[!t]
\hspace{3cm}
\epsfig{figure=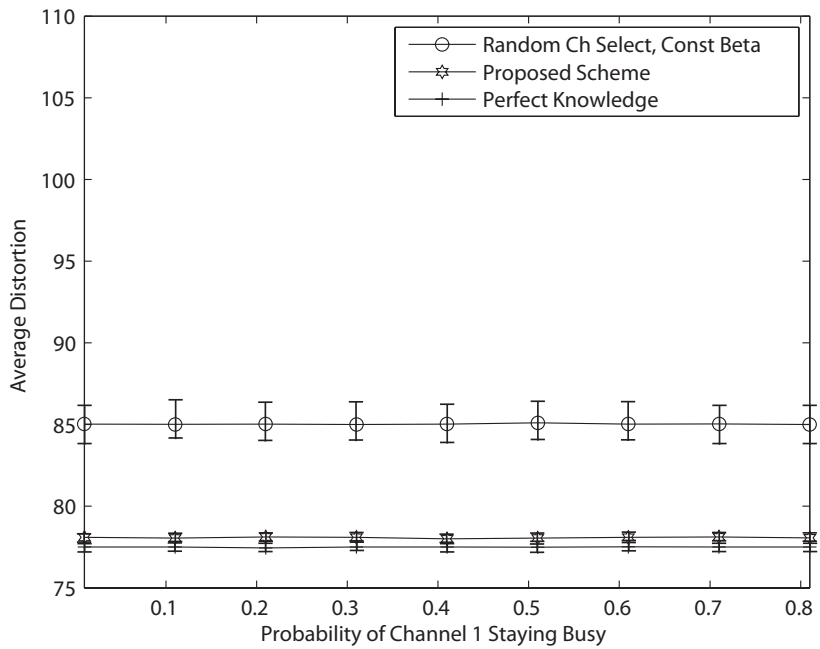, width=0.7\linewidth}
\caption{Two Channel Scenario: Average distortion vs. the probability of staying in the busy state of channel 1.}
\label{fig_2c_dist_stay_busy}
\end{figure}

In both cases, the SU utilization of our scheme is greater than the
non-adaptive scheme.  Our proposed scheme senses the surrounding environment to learn and
adapt channel selection.  However it takes several time slots for the
policy to learn the system state thus the performance of our scheme improves with
slower transition dynamics.  That is, our scheme approaches the
perfect case as $\mbox{Pr}\{X_{k+1}^1 = z | X_k^1 = v\}$ approaches
$0$ as is shown in Fig. \ref{fig_2c_spec_trans_busy}.  Our scheme provides closer to optimal performance when there is
a large difference in channel availability between the two channels as
it becomes easier to distinguish the better channel.  This is
demonstrated in Fig. \ref{fig_2c_spec_stay_busy} where the performance
of our scheme is more optimal at low $\mbox{Pr}\{X_{k+1}^1 = z | X_k^1
= z\}$ relative to  $\mbox{Pr}\{X_{k+1}^2 = z | X_k^2 = z\}$. In Fig. \ref{fig_2c_dist_stay_busy}, we show the average distortion for the probability channel $1$ stays busy. The average distortion of our scheme is better than the non-adaptive scheme.  Transition matrix parameters have little affect to the average distortion.  Our scheme outperforms the non-adaptive scheme because our scheme will select the channel with the better channel
gain and adapt the intra refreshing rate for the selected channel. 

\subsection{Effects of the Observation Matrix on Channel Selection Policy}
\begin{figure}[!t]
\hspace{3cm}
\epsfig{figure=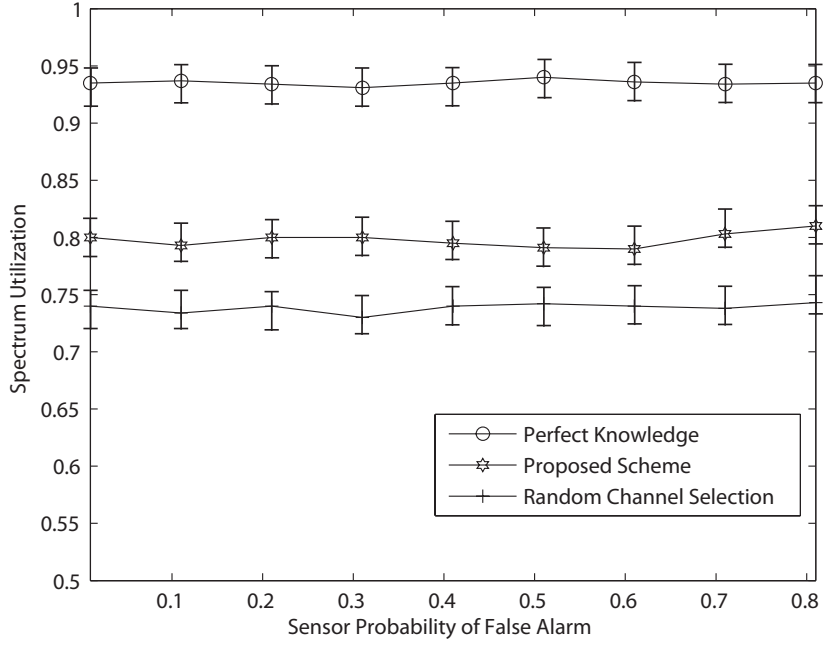, width=0.7\linewidth}
\caption{Two Channel Scenario: Spectrum utilization vs. the receiver channel estimation standard deviation, $\sigma$.}
\label{fig_2c_spec_epsilon}
\end{figure}

\begin{figure}[!t]
\hspace{3cm}
\epsfig{figure=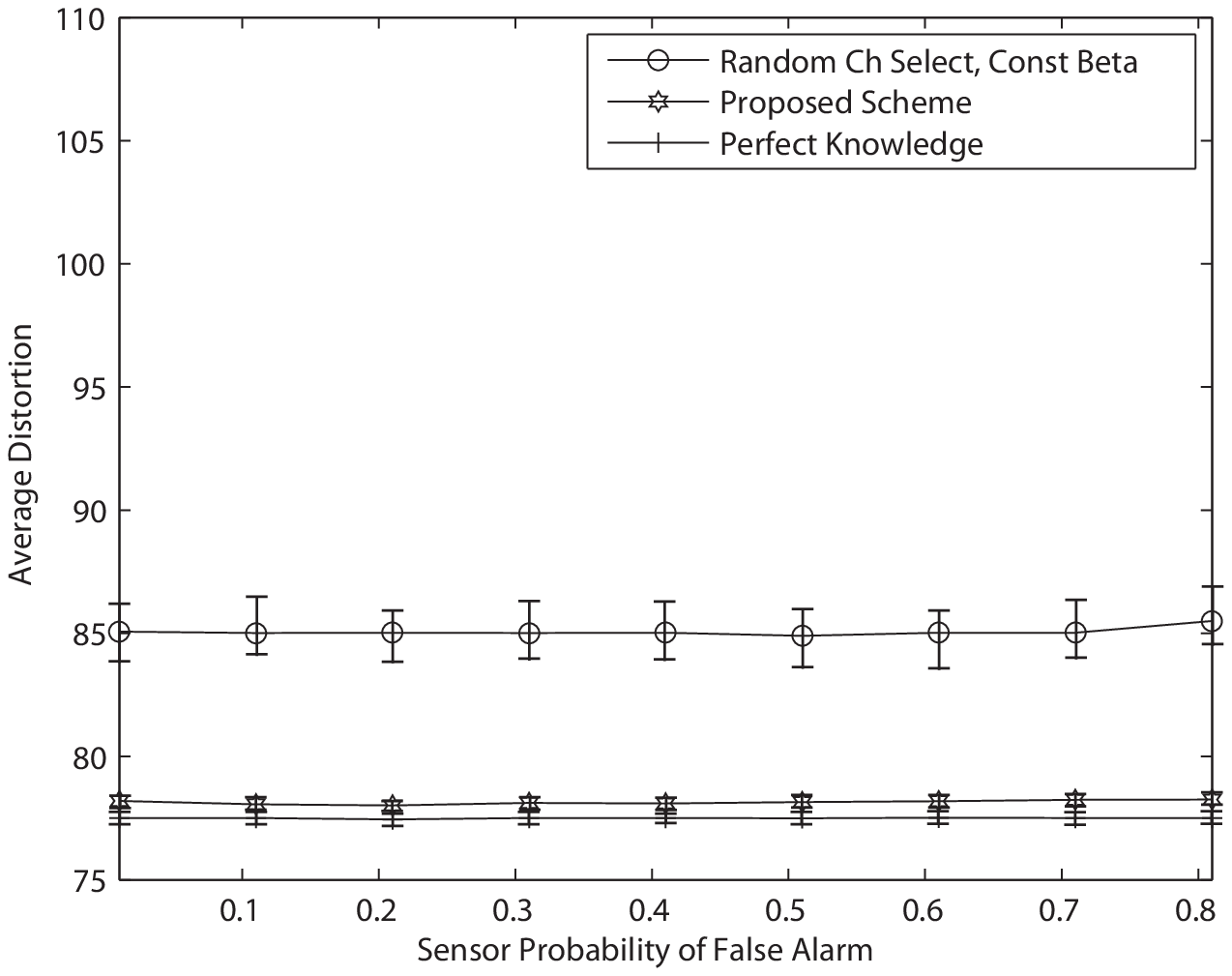, width=0.7\linewidth}
\caption{Two Channel Scenario: Average distortion vs. the receiver channel estimation standard deviation, $\sigma$.}
\label{fig_2c_dist_epsilon}
\end{figure}

SU with varying sensor operating point is shown in Fig. \ref{fig_2c_spec_epsilon}. $\mbox{Pr}\{X_{k+1}^1 = z | X_k^1 = z\} = 0.4$.
$\mbox{Pr}\{X_{k+1}^1 = z | X_k^1 = v\} = 0.15$. $\mbox{Pr}\{X_{k+1}^2
= z | X_k^2 = z\} = 0.6$. $\mbox{Pr}\{X_{k+1}^2 = z | X_k^2 = v\} =
0.2$.  Observation parameters are derived by operating characteristics
of the secondary users and are not likely to be different for each
channel.  Thus both channels are simulated with symmetrical observation parameters,
$\epsilon^1 = \epsilon^2 = \epsilon$ and $\sigma^1 = \sigma^2 =
\sigma$.  In Fig. \ref{fig_2c_spec_epsilon} we vary the spectrum
operating point $\epsilon$. In Fig. \ref{fig_2c_dist_epsilon} we show
the average distortion with varying $\epsilon$. The observation parameters are shown to have little affect on the SU
and average distortion performance of our proposed scheme.

These simulation results demonstrate some interesting trends in the
design and optimization of CR networks from a cross-layer design
perspective.  Adaptively adjusting the intra refreshing rate to
accommodate time varying wireless channels is an effective way to
reduce distortion.  By using all previous actions and observations we
can build an information state that becomes more accurate over time.
Performance of using the information state to select the intra refreshing
rate improves as the system dynamics slows down.  In a CR
environment the MAC access strategy is derived from the accuracy of
the spectrum sensor.  The total distortion is limited to the
availability of the channel.  Distortion performance will degrade if
primary usage increases or a very low system tolerable probability of
collision is required.  

\section{Conclusions and Future Work}
In this paper, we have presented an integrated approach for multimedia
transmission over cognitive radio networks. An important application layer parameter, intra refreshing rate, can be adjusted together with other parameters at other layers based on the sensed channel condition by the secondary users.  A low complexity dynamic
programming framework was
presented to obtain the optimal intra refreshing policy.  By modeling the system as a
Markov process, we have derived a POMDP for optimal channel selection
to minimize distortion while improving spectrum efficiency.
Simulation results demonstrated the performance gain by using the adaptive transmission scheme. Future work is in progress to consider other QoS at the application layer.

{\Large\textbf{Acknowledgment}}

We thank the reviewers for their detailed reviews and constructive comments, which have helped to improve the quality of this paper.

\begin{spacing}{0.5} 
\bibliographystyle{ieeetr}
\small{
\bibliography{D:/CA/Papers/Ref}}
\end{spacing} 

\newpage 

\end{document}